\documentclass{article}[11pts]
\usepackage[left=1in,right=1in,top=1in,bottom=1in]{geometry}

\usepackage[dvips]{graphicx}
\usepackage[latin1]{inputenc}
\usepackage{amssymb,amsmath,array}
\usepackage{booktabs}
\usepackage[table]{xcolor}

\usepackage{tabularray}
\usepackage[normalem]{ulem}
\usepackage{array,multirow,makecell}
\usepackage{amsmath}
\usepackage{amsfonts}
\usepackage{amssymb}
\usepackage[
	acronym,								
	xindy,									
	toc,									
  style=list					
	]
	{glossaries}	

\usepackage{graphicx}
\usepackage{xcolor}
\usepackage{placeins}
\usepackage{float}
\usepackage[boxed,ruled,lined ]{algorithm2e}
\usepackage[retainorgcmds]{IEEEtrantools}
\usepackage{hyperref}
\usepackage{authblk}
\usepackage{setspace}
\usepackage{bbold}
\usepackage{epstopdf}

\usepackage{stmaryrd}
\DeclareMathOperator*{\argminA}{arg\,min}

\usepackage[]{minitoc}

\newcommand*{\algotitle}[2]{%
  \stepcounter{algocf}%
  \hypertarget{algocf.title.\theHalgocf}{}%
  \NR@gettitle{#1}%
  \label{#2}%
  \addtocounter{algocf}{-1}%
}
\SetKwInput{Initialize}{Initialize  }
\SetKwInput{Iterate}{Iterate  }
\SetKwInput{Data}{Data }

\newcolumntype{R}[1]{>{\raggedleft\arraybackslash }b{#1}}
\newcolumntype{L}[1]{>{\raggedright\arraybackslash }b{#1}}
\newcolumntype{C}[1]{>{\centering\arraybackslash }b{#1}}

\usepackage{textcomp,marvosym}

\usepackage{nameref,hyperref}
\usepackage{array}

\begin{document}
\title{
Hierarchical novel class discovery for single-cell transcriptomic profiles
}

\author{Malek Senoussi$^1$ and Thierry Arti\`eres $^{1,2}$ and Paul Villoutreix$^{1,3}$
%
\thanks{Malek Senoussi et Paul Villoutreix were funded by the "Investissements d'Avenir" program of  the French government managed by the "Agence Nationale de la Recherche"(ANR-16-CONV-0001) and by the "Initiative d'Excellence d'Aix-Marseille Universit\'e - A*MIDEX".}
%
\vspace{.3cm}\\
%
1- Aix Marseille Univ, Universit\'e de Toulon, CNRS, LIS, \\Turing Centre for Living Systems, Marseille, France
%
\vspace{.1cm}\\
2- Ecole Centrale M\'editerran\'ee, Marseille, France
\vspace{.1cm}\\
3- Aix-Marseille Universit\'e, MMG, Inserm U1251, \\Turing Centre for Living systems,
Marseille, France\\
}


\maketitle

\begin{abstract}
One of the major challenge arising from single-cell transcriptomics experiments is the question of how to annotate the associated single-cell transcriptomic profiles. Because of the large size and the high dimensionality of the data, automated methods for annotation are needed. We focus here on datasets obtained in the context of developmental biology, where the differentiation process leads to a hierarchical structure.
We consider a frequent setting where both labeled and unlabeled data are available at training time, but the sets of the labels of labeled data on one side and of the unlabeled data on the other side, are disjoint. It is an instance of the Novel Class Discovery problem. The goal is to achieve two objectives, clustering the data and mapping the clusters with labels. We propose extensions of \textit{k-Means} and \textit{GMM} clustering methods for solving the problem and report comparative results on artificial and experimental transcriptomic datasets. Our approaches take advantage of the hierarchical nature of the data. 
\end{abstract}

\section{Introduction}
\begin{figure*}[h]
  \centering
  \includegraphics[width=\textwidth ]{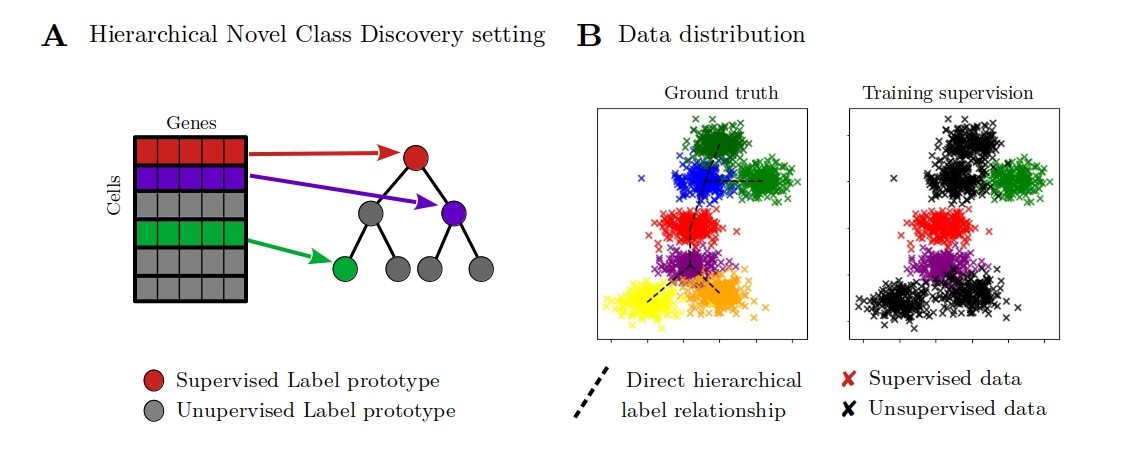}
\caption{Hierarchical Novel Class Discovery problem. A) The data under consideration are represented by a single-cell RNA sequencing (sc-RNA-Seq) matrix on the left, where each row represents a transcriptomic vector which belong to a class, where classes are organized in a hierarchy (lineage tree). A part of the data are labeled (colored according to their class) the other part is unlabeled (in gray). 
  B) The panel on the left shows the ground truth distribution of the data (one color per class). The right panel shows the available supervision for training the model, where only part of the labels include all supervised data, and the other part of labels include only unlabeled data, they are plotted in black.}

  \label{fig1}
\end{figure*}

Single cell RNA sequencing techniques (sc-RNASeq) generate large amounts of single cell transcriptomic profiles which are used to characterize cell states. These profiles need to be annotated into cell types. Their high dimensionality and quantity makes this task a difficult challenge \cite{ianevski2022fully, zhang2024cellstar, lyu2023cellann, pasquini2021automated, senoussi2024partial}. We focus here on a setting which occurs in many experimental contexts in single-cell transcriptomic studies: both labeled data and unlabeled data are available and one wants to infer the labels of the unlabeled data, but there is no overlap between the set of labels of the labeled data and the set of labels of unlabeled data, see Figure \ref{fig1}.  

We focus on studies of developmental biology, which have for object the process of development, i.e. how a single cell is turned into a multicellular organism. In this process, cells go through a differentiation process. At early stages of development, cells are undifferentiated and they become more specified later on. This differentiation process creates a branching structure, called the cell lineage tree \cite{wagner2020lineage}. Annotating single cell transcriptomic profiles in development requires to position these transcriptomic profiles into one of the cell type in the cell lineage tree. On Figure \ref{fig1} A, we show how this problem is turned into a computational one. The set of transcriptomic profiles form a large matrix where each row is a transcriptomic profile and each column is a gene. The aim is to associate each row to a node in the cell lineage tree. The specific task that we address here, is the one where some of the transcriptomic profiles can be positioned on the tree and some of the transcriptomic profiles have no annotation. In addition, because of the temporal continuity of the developmental process, we assume that the cell profiles vary almost continuously along the tree as illustrated on Figure \ref{fig1} B.


\section{Related works}

The problem we consider is an instance of what is known under the name Novel class discovery (\textit{NCD}) \cite{hsu2017learning, troisemaine2023novel}. The objective is to cluster an unlabeled dataset using guidance from a labeled dataset, with the strong constraint that there is no overlap between classes of labeled data (called hereafter supervised labels) and of unlabeled data (called hereafter unsupervised labels). The first formalization of the task seems to be the one by \cite{han2019learning} who used deep transfer clustering.  Recently a state-of-the-art approach named \textit{Autonovel} was proposed \cite{han21autonovel,zhao2021novel}. The method relies on a classification model which is classically trained in a supervised way with labeled data to discriminate between supervised labels. The model is decomposed into a feature extractor and a classifier. A second classifier is plugged on top of the feature extractor and aims at discriminating between the unsupervised labels. It is learned to output the same prediction (class) for unsupervised data that are considered similar, where similarity is inferred from the similarity of the encodings of unlabeled samples by the feature extractor. 

\textit{Zero Shot Learning} (\textit{ZSL}) is also quite related to our problem. \textit{ZSL} consists of learning to classify data in classes which were not seen at training time, similarly as what occurs in the task we consider \cite{tan2021survey}. It requires class-level semantic attributes  (e.g., word vectors \cite{socher2013zero}), and it relies on transferring knowledge from the training classes to the target classes. Some works have also studied hierarchical ZSL \cite{novack2023chils, schonfeld2019generalized}, but these works considered local hierarchical classification only.

Finally, the problem that we are addressing has some similarities with \textit{semi-supervised learning}. (\textit{SSL}) \cite{zhu2005semi}, and in particular with the \textit{Pseudo-labeling} strategy which is an instance of self-training \cite{amini2022self,Zhai2019ICCV} where the learning iterates a prediction step on unlabeled data, the addition of most confident predictions to the labeled training dataset, and a learning step with the completed training set 
\cite{fralick1967learning,triguero2015self,lee2013pseudo,grandvalet2004semi,oliver2018realistic}. Beyond mixture modeling, (e.g. \textit{GMMs}), which are natural probabilistic methods for semi-supervised learning \cite{miller1996mixture,nigam2000text}, many approaches relying on \textit{generative models} have been proposed \cite{yang2022survey}, \cite{radford2015unsupervised,salimans2016improved,kingma2014semi}, building on the success of Generative Adversarial Networks \cite{goodfellow2014generative} and Variational AutoEncoders \cite{kingma2013auto}, but these strategies do not fit well in our case since training data are often limited. 
Moreover contrary to our case in SSL, unlabeled and labeled data usually belong to the same labels. 

At the end, the task we tackle exhibits several specificities that prevent from applying standard methods. It is an instance of the \textit{NCD} problem, but where some additional knowledge (the hierarchical organization of classes) may be used in a zero-shot-learning like manner to enable inference in the hierarchical class taxonomy

\section{Method}

We first formalize the problem and propose to use a combined objective that consists of a standard clustering criterion and a label-cluster mapping criterion that we define from the prior knowledge of the hierarchical organization of the classes. We then detail two models that are learned to optimize this combined loss: a hierarchical \textit{k-means} based method (\textit{h-k-means}) which we extend to Gaussian Mixture Models for hierarchies (\textit{h-GMM}).

\subsection{Problem formalization}

We consider a labeled dataset $D_s = \{ (x_i,y_i) \in \mathbb{R}^d \times \mathcal{Y}_s ; i=1, ..., L \}$, whose samples belong to a set of, what we call, \textit{supervised} labels $\mathcal{Y}_s \subset \mathcal{Y}$, and an unlabeled dataset $D_u = \{ x_i \in \mathbb{R}^d ; i=L+1, ..., L+U \}$ whose samples belong to a set of, what we call, \textit{unsupervised} labels $\mathcal{Y}_u \subset \mathcal{Y}$, i.e. $\forall x_i \in D_u, y_i \in \mathcal{Y}_u$. As stated above we focus on the particular case where the sets of labels of the samples in $D_s$ and $D_u$ do not overlap, $\mathcal{Y}_s \cap \mathcal{Y}_u= \varnothing$. At last, we are given prior knowledge about the labels that consist of a hierarchical organization of these, i.e. the lineage tree.  We will note $\gamma$ a label in $\mathcal{Y}$, i.e. a node in the hierarchy, $f[\gamma]$ the father node of the node $\gamma$, and $\mathcal{A}({\gamma})$ the set of the ancestors of the node $\gamma$.

A quite naive solution would consist of first performing a clustering of the data in $D_u$ into $K$ clusters (with  $K=|\mathcal{Y}_u|$ the number of unsupervised labels) and second identifying a mapping $m$ between the $K$ clusters and the $K$ unsupervised labels in $\mathcal{Y}_u$. Then one could consider a classifier which operates as follows: it classifies all samples $x_i$ which belong to a cluster $k$ in the class $\gamma$ which is mapped to this cluster (i.e. $\gamma$ is such that $m(\gamma) = k$). The accuracy of this classifier could be written as:
\begin{equation}
    score (m, Z, \mathcal{Y}_u) =  \frac{1}{L+U}
    \sum_{i=L+1}^{L+U}  \mathbb{1}_{[m(z_i) = y_i]}
\end{equation}
where $\mathbb{1}_{[m(z_i) = y_i]}=1$ or $0$ depending of $m(z_i)$ being equal to $y_i$ or not, and $z_i \in \{1,...,K\}$ stands for the hidden variable associated with sample $x_i \in D_u$, i.e. the cluster $x_i$ belongs to. 

Unfortunately such an approach is hard to design as the identification problem (mapping the clusters with the unsupervised labels) is a difficult combinatorial problem: it is not straightforward to define an identification criterion nor to optimize it. 

\subsection{Hierarchical continuity loss}
We propose to use the known hierarchy of labels, the lineage, to designed clustering methods. We rely on a rather intuitive assumption, for transcriptomic data, which builds on this hierarchical organization. Indeed, one can assume that the dynamics of the expression of cells in the lineage tree (i.e. during the differentiation process) is continuous and slow, meaning a cell's gene expression should be close to the one of its mother/father cell \cite{wagner2020lineage, senoussi2024partial}. We now turn these assumptions into a loss term for optimzing the mapping.

Let us note $\mu_{\gamma} \in \mathbb{R}^d$ the representative vector, or mean, associated to label/node $\gamma$ in the hierarchy \footnote{We will call the representative vector of a node a\textit{ mean} vector to match the clustering literature although it might not be actually computed as the mean of the node's samples.}. Let note $f[\gamma]$ be the father class of a class $\gamma$ and $\epsilon_{\gamma} = \mu_{\gamma} - \mu_{f[\gamma]} $, the assumption states that 
$\forall \gamma, \Vert \epsilon_{\gamma}\Vert = \Vert \mu_{\gamma} -\mu_{f[\gamma]}\Vert $ is likely to be small. Moreover, applying the above decomposition ($\mu_{\gamma} = \mu_{f[\gamma]} + \epsilon_{\gamma} $) to all labels/nodes in the lineage tree yields, denoting $\mathcal{A}(\gamma)$ the set of ancestors in the lineage tree of a node $\gamma$, we get: 
\begin{equation}  
\forall {\gamma} , \mu_{\gamma}  =   \epsilon_{\gamma} + \sum\limits_{{\gamma}' \in \mathcal{A}({\gamma})} \epsilon_{{\gamma}'} 
\label{eq:defmu_epsilon}
\end{equation} 
And one can define what we call a \textit{continuity loss} term, $L_{cont}$, that one may want to minimize: 

\begin{align}
   L_{cont}:=& \sum\limits_{{\gamma} \in \mathcal{Y}} \Vert \mu_{\gamma} -\mu_{f[{\gamma}]}\Vert^2 \\
            &= \sum\limits_{{\gamma} \in \mathcal{Y}} \Vert \epsilon_{\gamma} \Vert^2 
\label{eq:GlobalLossEpsilon}            
\end{align}

The above criterion is differentiable with respect to the $\mu$ vectors (and w.r.t. the $\epsilon$ vectors) and may be used as an additional loss term to learn the parameters of a hierarchical \textit{k-means} or \textit{GMM} approach, as we will detail in the following. 


\subsection{Hierarchical \textit{k-means} (\textit{h-k-means})}
We present a method which is an extension of the \textit{k-means} clustering algorithm to the hierarchal classification case we are interested in, it relies on the continuity loss term (in Eq. (\ref{eq:GlobalLossEpsilon})). The hierarchical \textit{k-means} (\textit{h-k-means}) model we want to learn is a classical \textit{k-means} performed on $D_u$ samples, with an additional continuity loss term that is expected to help guide the mapping of clusters to the correct labels in the hierarchy. The objective function is written as:
\begin{align} 
\argminA_{\{\epsilon_{\gamma}, \gamma \in \mathcal{Y}_u\}, \{\tilde{y}_i, i=L+1..L+U\}} &\sum\limits_{x_i \in D_{s}} ||x_i - \mu_{y_i} ||^2 \nonumber\\
&+ \lambda_u  \sum\limits_{x_i \in D_{u}} ||x_i - \mu_{\tilde{y}_i} ||^2  + \lambda_{\epsilon} \sum\limits_{\gamma \in \mathcal{Y}} || \epsilon_{\gamma} ||^2
\label{eq:hKmeansObjective}
 \end{align}
where all $\mu$'s are defined as sums of $\epsilon$ vectors according to Eq. (\ref{eq:defmu_epsilon}), and where $\tilde{y}_i$ are latent variables that need to be optimized.

To solve this problem, we propose to use, as for standard \textit{k-means}, an iterative hard assignment EM-like algorithm which first updates the pseudo-labeling $\tilde{y}_i$ 
 of the unsupervised data as in the standard \textit{k-means} algorithm based on current parameters, according to 
$\tilde{y_i}  = \argminA_{\gamma \in \mathcal{Y}_u }   ||x_i - \mu_{\gamma} ||$ 
and then reestimate the representative of all unsupervised labels $(\mu_{\gamma})_{\gamma \in \mathcal{Y}_u}$ by minimizing the objective function in Eq. (\ref{eq:hKmeansObjective}) with respect to $\varepsilon$ vectors using gradient descent, while $\tilde{y_i}$ remain fixed. The hyperparameters $\lambda_u ,\lambda_{\varepsilon}$ are set up by cross validation.

\subsection{Hierarchical Gaussian Mixture Model (\textit{h-GMM})}

We additionally considered Hierarchical Gaussian Mixture Models (\textit{h-GMMs}) which extend hierarchical \textit{k-means} alike \textit{GMMs} extend \textit{k-means}. We assume that data from $D_u$ follow a \textit{GMM} distribution, with one Gaussian component per cluster/label/node. 

Like with \textit{h-k-means}, we encode the hierarchical information through the expression of the mean of each Gaussian component on the mixture:
\begin{equation}  
\forall \gamma, \mu_{\gamma} = \epsilon_{\gamma} + \sum\limits_{\gamma' \in \mathcal{A}(\gamma)} \epsilon_{\gamma'}       
\end{equation} 

We note the parameters as $\Theta = \{ (\pi_{\gamma}, \mu_{\gamma}, \Sigma_{\gamma}), \gamma \in \mathcal{Y}_u  \}$, or equivalently $\Theta = \{ (\pi_{\gamma}, \epsilon_{\gamma}, \Sigma_{\gamma}), \gamma \in \mathcal{Y}_u \}$. The optimization is performed through the maximization of the log-likelihood of the data. One can optimize the penalized conditional expected likelihood of the complete data $Q(\Theta, \Theta^{(t)})$, given the old values for parameters $\Theta^{(t)})$. The addition of the hierarchical regularization can be done directly in the auxiliary function $Q$, \cite{he2010laplacian}, leading to:
\begin{equation}
  \Tilde{Q}(\Theta, \Theta^{t}) = \sum\limits_{i=1}^{L+U} \sum\limits_{\gamma \in \mathcal{Y}}  p(z=\gamma |x_i, \Theta^{t} )log p(x_i,\gamma | \Theta ) + log \pi_{\gamma}   - \lambda_{\epsilon} \sum\limits_{\gamma \in \mathcal{Y}} || \epsilon_{\gamma} ||^2
  \label{eq:Qfunction_hGMM}
\end{equation}
where $p(z=\gamma |x_i, \Theta^{t} )=0$ if $(x_i,\gamma) \in D_u \times \mathcal{Y}_s $ or if $(x_i,\gamma)\in D_s \times \mathcal{Y} \setminus \{y_i\} $ to take into account the available supervision on data in $D_s$ and to enforce that unsupervised data should be labeled in $\mathcal{Y}_u$.

We use a regularized Expectation Maximization algorithm to learn \textit{h-GMMs}. 
This algorithm is close to the standard EM algorithm, the difference lies in the \textit{M-step} and concerns the optimization of the means. While the  $\{(\pi_{\gamma}, \Sigma_{\gamma}), \gamma \in \mathcal{Y}_u\}$ parameters are optimized similarly as for standard \textit{GMMs}, there is no closed form solution for the means $\{\mu_{\gamma}, \gamma \in \mathcal{Y}_u\}$ and the $\varepsilon$ vectors. 
We rather perform a few steps of gradient ascent (\cite{gepperth2021gradient}) every M-step to maximize $Q(\Theta, \Theta^{t})$ with respect to $\{\epsilon_{\gamma}, \gamma \in \mathcal{Y}_u\}$.

\section{Datasets}

We evaluated the methods on artificial and experimental datasets. We used three artificial datasets generated with the library Prosstt (named Branches-7, Binary-4, Binary-6 in the following) \cite{papadopoulos2019prosstt}, and two experimental scRNA-Seq datasets from \textit{Myeloid Progenitor} \cite{paul2015transcriptional} and \textit{S. Mediteranea}  \cite{plass2018cell}. 
The main characteristics of the five datasets are summarized in Table \ref{tab:c5t0}.

\begin{table}[h]
    \centering
    \begin{tabular}{c||llccc}
    \hline  &   Branches-7&Binary-4&Binary-6& \textit{Myeloid Progenitor}& \textit{S. mediterranea}\\
    \hline \#features  &   50&50&50 & 100&50\\
     \#labels  &   36&31&127 & 26& 51\\
     \#samples  &  2800&1240&5000& 2730& 7595\\
     \hline
    \end{tabular}
    \caption{Simulated and experimental datasets main characteristics} 
    \label{tab:c5t0}
\end{table}

For all the datasets, the prior knowledge is the hierarchical organisation of the labels, which is given as a tree of labels. We will note $H \in \mathbb{R}^{C,C}$ the matrix of distances (defined as the shortest path in the tree) between pairs of labels, where $C$ is the number of labels.

\subsection{Simulated Datasets}

To simulate realistic datasets of transcriptomic profiles incorporating the hierarchical structure observed in development, we used the \textit{prosstt} library \cite{papadopoulos2019prosstt}. This simulation library offers the ability to generate differentiation trajectories with various topologies. Following \cite{senoussi2024partial}, we focused particularly on two scenarios. Both of them are structured with a tree topology. First, we considered the \textit{Branches} case where the tree is asymmetric and each bifurcation leads to two types of branches, one that follow a linear topology and the other one goes on dividing. The number of labels is $\sum_{i=0}^{t+1} i$ where $t$ is the depth of the tree. Second, we considered the \textit{Binary} case corresponding to the situation where, at each bifurcation, the tree divide into two binary trees. The number of labels is given by $\sum_{i=0}^{t+1} 2^i$ where $t$ is the depth of the tree.


Once a topology is set, the generation of a dataset depends on two parameters that govern the dynamics of gene expression along the topology. 
The two parameters are the \textit{number of genetic programs}, denoted as $g$, and the \textit{noise level}, noted $\alpha$. The dynamics of gene expression across the differentiation trajectories are modeled using genetic programs that are instantiated as random walks on a tree. Using a large number of genetic programs increases the combinatorics of gene expression in each branch, which leads to larger differences between data in various parts of the tree, hence easier classification or clustering. The noise level $\alpha$ controls the characteristics of the negative binomial probability distribution that is used to generate realistic gene expression \cite{papadopoulos2019prosstt}. We used $\alpha=0.1$ and $g=50$. 

\subsection{Experimental Datasets} \label{data}



To explore our models on real experimental datasets, we turned to classical datasets of developmental biology also exploited in \cite{senoussi2024partial}. The first one is an atlas of development of the planaria \textit{S. mediterranea} at single cell resolution. We considered the lineage tree obtained in \cite{plass2018cell} using the PAGA method \cite{wolf2019paga}. The dataset consists of 21612 transcriptomic vectors of dimension 28065, distributed among 51 different labels. 
The labels represent a collection of cell types arranged as a tree based on cell differentiation. The tree is not deep as most branches contain between two and three labels, the longest contains six labels. To obtain a balanced number of transcriptomic profiles associated to each labels, we subsampled the initial dataset. The dataset that we used ended up containing 7595 transcriptomic profiles.
 
The second experimental dataset, \textit{Myeloid progenitors}, is an atlas of hematopoiesis at single-cell resolution \cite{paul2015transcriptional}. The 2730 transcriptomic profiles are composed of 10719 genes distributed among 26 labels. The 26 labels correspond to the various cell types observed in blood cells differentiation \cite{weiskopf2017myeloid,spangrude1988purification}.


For computing efficiency, as it is done usually, dimensionality reduction using PCA was applied to the two datasets. We kept the first 50 and 100 components for \textit{S. mediterranea} and \textit{Myeloid progenitors}, respectively. 

 



\section{Experiments}


We investigated the behavior of the methods in various settings, with only a few unsupervised labels up to a large percentage of such labels in $\mathcal{Y}$. The proportion of unsupervised data, $ | \mathcal{Y}_u | / |\mathcal{Y}|$, takes values of 0.1, 0.25, and 0.5. For a given percentage value, we create the appropriate datasets by first random sampling of labels in $\mathcal{Y}_s$, $\mathcal{Y}_u$ according to the desired proportion of unsupervised data, with no overlap between the two sets, then we create $X_s$ and $X_u$. 
For each setting (dataset, percentage of unsupervised labels, split between $\mathcal{Y}_s$ and $\mathcal{Y}_u$), we split the data in train-test sets for both supervised $X_s = X_s^{train} \cup X_s^{test}$ and unsupervised datasets $X_u = X_u^{train} \cup X_u^{test}$ with a proportion of 0.2 for the test size. The performances are measured on the $X_u^{test}$ set.
The experiments are repeated 5 times per setting and reported performances are averaged results. Note that this setup is sensitive to high variation, especially with a small number of labels. Moreover, the data are uniformly distributed among the labels in the artificial datasets, but not in the experimental ones.

The code for the various methods is available at \url{https://github.com/MalekSnous/hNCD-scRNAseq}.

\subsection{Metrics}

We use both clustering (ACC) and classification  (micro-f1) metrics. 
The Accuracy Cluster Classification (\textit{ACC}) is standard practice for evaluating clustering results \cite{vaze2022generalized}. Given a ground truth labelling of samples into classes and a labelling output by a clustering algorithm, the \textit{ACC} metric relies on a mapping of clusters onto classes, where every sample from a given cluster is classified as the class the cluster is mapped to. The \textit{ACC} is defined as the maximum accuracy one can reach with any possible mapping. Finding the optimal mapping, the one maximizing the accuracy, is not straightforward because it requires solving a combinatorial problem but a sub-optimal mapping may efficiently be obtained using the Hungarian algorithm. 

We report and compare experimental results obtained with several methods: Non-hierarchical clustering methods: \textit{k-}Means, \textit{GMMs}, a novel class discovery state of the art model (AutoNovel) and our hierarchical methods \textit{h-k-Means} and \textit{h-GMMs}.

\subsection{Results}

We start with the analysis of clustering results alone, see Table \ref{tab:acc-comparison}. The first comment concerns the variance of the results which is quite large, especially when the number of unsupervised labels is small. This comes from the fact that if only two or three classes are unsupervised (the classes in $\mathcal{Y}_u$), the data belonging to these classes may be (randomly) drawn from very far to very close. For the same reason while the superiority of methods over others appears significant in some cases, it is not always the case, despite large differences in averaged performances. For instance in the first row of Table \ref{tab:acc-comparison}, the hierarchical models look, but are not, significantly better than non-hierarchical methods. Next, one may see that all methods perform quite well on the clustering tasks whatever the setting, for artificial as well as for experimental datasets. The results obtained with half unsupervised labels, meaning up to 64 unsupervised labels on artificial datasets (around $0.36$ ACC for \textit{GMMs}) and up to 25 unsupervised labels for experimental datasets (around $0.50$ ACC for \textit{GMMs}) show that the data are likely not too complex to discriminate. Moreover standard clustering methods are quite competitive with the state-of-the-art \textit{Autonovel} baseline for the Novel Class Dicscovery problem.

Finally and importantly, one sees that hierarchical methods perform significantly better than their non-hierarchical counterparts in many settings. This was expected but not straightforward as the addition of a hierarchical regularization in the optimization of the models could have reduced the pure clustering performance. For example, for Binary-6 with 12 unsupervised labels, the ACC of \textit{h-k-Means} is at 0.65, while the corresponding \textit{k -means} is 0.50, and that of \textit{h-GMM} is 0.64, while\textit{ GMM} is 0.55.

\begin{table}[h]
    \centering
    \begin{tabular}{lc||cc|ccc} 
        \toprule
           ACC&&  \multicolumn{5}{c}{Methods}\\ 
  Dataset& & \multicolumn{2}{c}{Hierarchical models}& \multicolumn{3}{c}{Non hierarchical models}\\ 
  $|\mathcal{Y}|$&  $|\mathcal{Y}_u|$ &  \textit{ h-k-means}& \textit{h-GMM}& \textit{k -means}&\textit{ GMM}&\textit{Autonovel}\\ 
 \midrule
     \textit{Branches-7}&3&\underline{0.90 $\pm$ 0.05} & \underline{0.90 $\pm$ 0.05}  & 0.64 $\pm$ 0.1 & 0.64 $\pm$ 0.10& 0.80 $\pm$ 0.10\\ 
            36&9& \underline{0.62 $\pm$ 0.03 }& 0.64 $\pm$ 0.05 & 0.52 $\pm$ 0.02 & 0.60 $\pm$ 0.03 & 0.63 $\pm$ 0.04 \\ 
             &18& 0.44 $\pm$ 0.11 & 0.44 $\pm$ 0.09 & 0.43 $\pm$ 0.04 & 0.47 $\pm$ 0.04 &  \textbf{0.54 $\pm$ 0.03 } \\ 
 \midrule           
        \textit{Binary-4}&3& 0.84 $\pm$ 0.13 & 0.84 $\pm$ 0.13 & 0.76 $\pm$ 0.12 & 0.73 $\pm$ 0.15 &  0.84 $\pm$ 0.03 \\ 
           31&7& 0.64 $\pm$ 0.05 & 0.69 $\pm$ 0.02 & 0.57 $\pm$ 0.07 & 0.67 $\pm$ 0.07 &   0.65 $\pm$ 0.04\\ 
           &15& 0.47 $\pm$ 0.07 & 0.44 $\pm$ 0.08 & 0.43 $\pm$ 0.02 & 0.49 $\pm$ 0.04 & 0.50 $\pm$ 0.05\\ 
 \midrule
    \textit{Binary-6}& 12 &  \underline{0.65 $\pm$ 0.08} & 0.64 $\pm$ 0.09 & 0.50 $\pm$ 0.07& 0.55 $\pm$ 0.10&    0.65 $\pm$ 0.08\\ 
      127   & 32 & \underline{0.44 $\pm$ 0.01} & 0.44 $\pm$ 0.01 & 0.37 $\pm$ 0.02 & 0.45 $\pm$ 0.04 &  0.30 $\pm$ 0.03\\
           & 64  & \underline{0.31 $\pm$ 0.01} & 0.31 $\pm$ 0.01 & 0.29 $\pm$ 0.01& \underline{\textbf{0.36 $\pm$ 0.01}} &   0.14 $\pm$ 0.01\\ 

\midrule
    
        \textit{Myeloid}&2& 0.93 $\pm$ 0.13 & 0.93 $\pm$ 0.13 & 0.92 $\pm$ 0.14 & 0.92 $\pm$ 0.14 &  0.88 $\pm$ 0.18  \\ 
           \textit{Progenitor}&6&0.71 $\pm$ 0.18 & 0.78 $\pm$ 0.19 & 0.75 $\pm$ 0.14 & 0.72 $\pm$ 0.12 & 0.60 $\pm$ 0.06\\ 
           26&13&  0.67 $\pm$ 0.06 & 0.66 $\pm$ 0.05 & 0.64 $\pm$ 0.03 & 0.66 $\pm$ 0.06 &  0.59 $\pm$ 0.06  \\ 
 
 \midrule
           
        \textit{S. mediterranea}&5&   \underline{0.64 $\pm$ 0.10}& 0.63 $\pm$ 0.11 & 0.47 $\pm$ 0.10& 0.62 $\pm$ 0.09 &  0.62 $\pm$ 0.05\\ 
           &12&0.56 $\pm$ 0.13 & 0.56 $\pm$ 0.12 & 0.49 $\pm$ 0.09 & 0.53 $\pm$ 0.10& 0.55 $\pm$ 0.04\\ 
           51&25&   0.36 $\pm$ 0.11 & 0.37 $\pm$ 0.09 & 0.42 $\pm$ 0.02 & \underline{\textbf{0.49 $\pm$ 0.02}} &  0.30 $\pm$ 0.03\\
           \bottomrule
    \end{tabular}
    \vspace{0.2cm}
    \caption{ Clustering results (ACC metric) for hierarchical methods (\textit{h-k-Means} and \textit{h-GMM}), clustering methods (\textit{k-Means} and \textit{GMM}) and novel class discovery (\textit{Autonovel}): A \textbf{result in bold} means that the corresponding method significantly outperforms the other methods, according to a paired t-test with p-value $< 0.1$. An \underline{underlined result} indicates a significative difference between the hierarchical method and its non-hierarchical counterpart (e.g. \textit{h-k-means} vs \textit{k-means}) according to a paired t-test with p-value $< 0.1$. 
    }
    \label{tab:acc-comparison}
\end{table}

We now turn to the classification results reported in Table \ref{tab:f1-comparison}. Accuracies are reported for the two hierarchical methods (\textit{h-k-Means} and \textit{h-GMM}). One key element to assess the relevance of the methods lies in comparing the \textit{f1-score} performance of hierarchical methods in Table \ref{tab:f1-comparison} and the \textit{ACC} performance in Table \ref{tab:acc-comparison}. The ACC performance may be viewed as an upper bound of the achievable performance by hierarchical methods. Reaching a f1-score close to the ACC performance is then very promising. The two hierarchical methods do perform well on artificial datasets. Very often the f1-score is not far from the ACC performance. For instance, with Binary-4 dataset, \textit{h-k-Means} reaches an accuracy of 0.62 (for 7 unsupervised labels) while its ACC was 0.64 (see Tab.\ref{tab:acc-comparison}). With Branches-7 \textit{h-GMM} reaches 0.61 f1-score while its ACC was 0.64. 

When comparing the ACC performance of a hierarchical model and its f1 performance it may be seen that for two of the datasets (Branches-7 and Binary-4) the performance of a hierarchical method is not significantly lower than its ACC performance, which is in itself a promising result (see Table\ref{tab:acc-comparison} caption).

Finally, it is worth noting that hierarchical methods may reach a higher accuracy than the ACC performance of clustering methods, \textit{k-Means} and \textit{GMM} (see Table \ref{tab:acc-comparison}). This is particularly the case in the easier cases only (simulated datasets, few unsupervised labels) but still shows the potential power of these approaches.


At the end, it looks like artificial datasets offer favourable settings for hierarchical methods. This is in line with other results \cite{senoussi2024partial} where exploiting the hierarchy might bring improvement but not in a systematic way. This might come from the nature of the data. While our assumption on the distribution of data in the nodes of the hierarchy is obviously satisfied for simulated data, it is not clear how this assumption holds in experimental datasets as the differentiation process might not always lead to a hierarchy and the hierarchical prior is usually established manually and might therefore be less reliable \cite{wagner2020lineage}.

\begin{table}[h]
    \centering
    \begin{tabular}{l|ccc} 
        \toprule
           f1&  \multicolumn{3}{c}{ }\\ 
  Dataset& & \multicolumn{2}{c}{Hierarchical methods}\\ 
  $|\mathcal{Y}|$&  $|\mathcal{Y}_u|$ &  \textit{ h-k-means}& \textit{h-GMM}\\ 
 \midrule
    \textit{Branches-7}&3& 0.71 $\pm$ 0.25&  0.83 $\pm$ 0.19  \\ 
           36&9& 0.60 $\pm$ 0.04 & 0.61 $\pm$ 0.05  \\ 
           &18& \dotuline{0.29 $\pm$ 0.05}& \dotuline{0.29 $\pm$ 0.06 } \\ 
   \midrule         
    \textit{Binary-4}&3& 0.84 $\pm$ 0.16& 0.84 $\pm$ 0.16\\ 
           31&7& 0.62 $\pm$ 0.09 &  0.58 $\pm$ 0.09  \\ 
           &15& \dotuline{0.36 $\pm$ 0.05} &  0.36 $\pm$ 0.07   \\ 
 \midrule   
   \textit{Binary-6}&12& \dotuline{0.58 $\pm$ 0.05} & \dotuline{0.57 $\pm$ 0.05}\\ 
    127    &32& \dotuline{0.35 $\pm$ 0.02}& \dotuline{0.35 $\pm$ 0.02}      \\ 
           &64& \dotuline{0.19 $\pm$ 0.01}  & \dotuline{0.19 $\pm$ 0.01}   \\ 
 \midrule    
    \textit{Myeloid}&2&  0.64 $\pm$ 0.36& 0.59 $\pm$ 0.38 \\ 
    \textit{Progenitor}&6&  \dotuline{0.09 $\pm$ 0.16} & \dotuline{0.18 $\pm$ 0.16} \\ 
           26&13&  \dotuline{ 0.03 $\pm$ 0.03}  & \dotuline{0.14 $\pm$ 0.09}  \\ 
 \midrule      
     \textit{S. mediterranea}&5&     \dotuline{0.26 $\pm$ 0.11} &  \dotuline{0.16 $\pm$ 0.13 }\\ 
           51&12& \dotuline{0.06 $\pm$ 0.03 } & \dotuline{0.08 $\pm$ 0.04 }\\ 
           &25&   \dotuline{0.03 $\pm$ 0.02} & \dotuline{0.03 $\pm$ 0.01 } \\
           \bottomrule

    \end{tabular}
    
    \vspace{0.2cm}
    
    \caption{Hierarchical classification results (f1 metric) for hierarchical methods: an \dotuline{dotulined result} means that the ACC performance of the hierarchical model is significantly better than its f1 performance (according to a paired t-test with p-value $<0.1$) while a non-dotulined result indicates no significant difference. }
    \label{tab:f1-comparison}
\end{table}

\section{Discussion}

We proposed and compared several methods for a Novel Class Discovery task where the labels have a hierachical structure.
This is a preliminary work we have done for dealing with a recurrent feature of transcriptomics datasets, when part of the training data are labeled but with a subset only of the set of labels in the lineage hierarchy, and part of the data are unlabeled but are known to belong to another set of labels. 
Our results show that the hierarchical methods we propose do outperform clustering methods with respect to pure clustering metrics, indicating that taking into account the hierarchy  does help the clustering of the data. Moreover, we show that the hierarchical methods reach performances which are often close to an empirically estimated upper bound of their performance. 

It looks like the methods presented here may reach high performances in favourable and simpler settings, when the hierarchical feature is strongly present in the data and when the percentage of unsupervised labels is moderate, but are less efficient when the percent stage is large and on experimental datasets whose hierarchical feature is less obvious, this will be the objective of our future research to improve the methods in more difficult settings.







\bibliographystyle{unsrt}
\bibliography{biblio}

\begin{thebibliography}{10}

\bibitem{ianevski2022fully}
Aleksandr Ianevski, Anil~K Giri, and Tero Aittokallio.
\newblock Fully-automated and ultra-fast cell-type identification using specific marker combinations from single-cell transcriptomic data.
\newblock {\em Nature communications}, 13(1):1246, 2022.

\bibitem{zhang2024cellstar}
Ying Zhang, Huaicheng Sun, Wei Zhang, Tingting Fu, Shijie Huang, Minjie Mou, Jinsong Zhang, Jianqing Gao, Yichao Ge, Qingxia Yang, et~al.
\newblock Cellstar: a comprehensive resource for single-cell transcriptomic annotation.
\newblock {\em Nucleic Acids Research}, 52(D1):D859--D870, 2024.

\bibitem{lyu2023cellann}
Pin Lyu, Yijie Zhai, Taibo Li, and Jiang Qian.
\newblock Cellann: a comprehensive, super-fast, and user-friendly single-cell annotation web server.
\newblock {\em Bioinformatics}, 39(9):btad521, 2023.

\bibitem{pasquini2021automated}
Giovanni Pasquini, Jesus Eduardo~Rojo Arias, Patrick Sch{\"a}fer, and Volker Busskamp.
\newblock Automated methods for cell type annotation on scrna-seq data.
\newblock {\em Computational and Structural Biotechnology Journal}, 19:961--969, 2021.

\bibitem{senoussi2024partial}
Malek Senoussi, Thierry Artieres, and Paul Villoutreix.
\newblock Partial label learning for automated classification of single-cell transcriptomic profiles.
\newblock {\em PLoS Computational Biology}, 20(4):e1012006, 2024.

\bibitem{wagner2020lineage}
Daniel~E Wagner and Allon~M Klein.
\newblock Lineage tracing meets single-cell omics: opportunities and challenges.
\newblock {\em Nature Reviews Genetics}, 21(7):410--427, 2020.

\bibitem{hsu2017learning}
Yen-Chang Hsu, Zhaoyang Lv, and Zsolt Kira.
\newblock Learning to cluster in order to transfer across domains and tasks.
\newblock {\em arXiv preprint arXiv:1711.10125}, 2017.

\bibitem{troisemaine2023novel}
Colin Troisemaine, Vincent Lemaire, St{\'e}phane Gosselin, Alexandre Reiffers-Masson, Joachim Flocon-Cholet, and Sandrine Vaton.
\newblock Novel class discovery: an introduction and key concepts.
\newblock {\em arXiv preprint arXiv:2302.12028}, 2023.

\bibitem{han2019learning}
Kai Han, Andrea Vedaldi, and Andrew Zisserman.
\newblock Learning to discover novel visual categories via deep transfer clustering.
\newblock In {\em Proceedings of the IEEE/CVF International Conference on Computer Vision}, pages 8401--8409, 2019.

\bibitem{han21autonovel}
Kai Han, Sylvestre-Alvise Rebuffi, Sebastien Ehrhardt, Andrea Vedaldi, and Andrew Zisserman.
\newblock Autonovel: Automatically discovering and learning novel visual categories.
\newblock {\em IEEE Transactions on Pattern Analysis and Machine Intelligence (TPAMI)}, 2021.

\bibitem{zhao2021novel}
Bingchen Zhao and Kai Han.
\newblock Novel visual category discovery with dual ranking statistics and mutual knowledge distillation.
\newblock {\em Advances in Neural Information Processing Systems}, 34:22982--22994, 2021.

\bibitem{tan2021survey}
Chufeng Tan, Xing Xu, and Fumin Shen.
\newblock A survey of zero shot detection: methods and applications.
\newblock {\em Cognitive Robotics}, 1:159--167, 2021.

\bibitem{socher2013zero}
Richard Socher, Milind Ganjoo, Christopher~D Manning, and Andrew Ng.
\newblock Zero-shot learning through cross-modal transfer.
\newblock {\em Advances in neural information processing systems}, 26, 2013.

\bibitem{novack2023chils}
Zachary Novack, Julian McAuley, Zachary~Chase Lipton, and Saurabh Garg.
\newblock Chils: Zero-shot image classification with hierarchical label sets.
\newblock In {\em International Conference on Machine Learning}, pages 26342--26362. PMLR, 2023.

\bibitem{schonfeld2019generalized}
Edgar Schonfeld, Sayna Ebrahimi, Samarth Sinha, Trevor Darrell, and Zeynep Akata.
\newblock Generalized zero-and few-shot learning via aligned variational autoencoders.
\newblock In {\em Proceedings of the IEEE/CVF conference on computer vision and pattern recognition}, pages 8247--8255, 2019.

\bibitem{zhu2005semi}
Xiaojin~Jerry Zhu.
\newblock Semi-supervised learning literature survey.
\newblock 2005.

\bibitem{amini2022self}
Massih-Reza Amini, Vasilii Feofanov, Loic Pauletto, Emilie Devijver, and Yury Maximov.
\newblock Self-training: A survey.
\newblock {\em arXiv preprint arXiv:2202.12040}, 2022.

\bibitem{Zhai2019ICCV}
Xiaohua Zhai, Avital Oliver, Alexander Kolesnikov, and Lucas Beyer.
\newblock S4l: Self-supervised semi-supervised learning.
\newblock In {\em Proceedings of the IEEE/CVF International Conference on Computer Vision (ICCV)}, 11 2019.

\bibitem{fralick1967learning}
S~Fralick.
\newblock Learning to recognize patterns without a teacher.
\newblock {\em IEEE Transactions on Information Theory}, 13(1):57--64, 1967.

\bibitem{triguero2015self}
Isaac Triguero, Salvador Garcia, and Francisco Herrera.
\newblock Self-labeled techniques for semi-supervised learning: taxonomy, software and empirical study.
\newblock {\em Knowledge and Information systems}, 42:245--284, 2015.

\bibitem{lee2013pseudo}
Dong-Hyun Lee et~al.
\newblock Pseudo-label: The simple and efficient semi-supervised learning method for deep neural networks.
\newblock In {\em Workshop on challenges in representation learning, ICML}, volume~3, page 896. Atlanta, 2013.

\bibitem{grandvalet2004semi}
Yves Grandvalet and Yoshua Bengio.
\newblock Semi-supervised learning by entropy minimization.
\newblock {\em Advances in neural information processing systems}, 17, 2004.

\bibitem{oliver2018realistic}
Avital Oliver, Augustus Odena, Colin~A Raffel, Ekin~Dogus Cubuk, and Ian Goodfellow.
\newblock Realistic evaluation of deep semi-supervised learning algorithms.
\newblock {\em Advances in neural information processing systems}, 31, 2018.

\bibitem{miller1996mixture}
David~J Miller and Hasan Uyar.
\newblock A mixture of experts classifier with learning based on both labelled and unlabelled data.
\newblock {\em Advances in neural information processing systems}, 9, 1996.

\bibitem{nigam2000text}
Kamal Nigam, Andrew~Kachites McCallum, Sebastian Thrun, and Tom Mitchell.
\newblock Text classification from labeled and unlabeled documents using em.
\newblock {\em Machine learning}, 39:103--134, 2000.

\bibitem{yang2022survey}
Xiangli Yang, Zixing Song, Irwin King, and Zenglin Xu.
\newblock A survey on deep semi-supervised learning.
\newblock {\em IEEE Transactions on Knowledge and Data Engineering}, 35(9):8934--8954, 2022.

\bibitem{radford2015unsupervised}
Alec Radford, Luke Metz, and Soumith Chintala.
\newblock Unsupervised representation learning with deep convolutional generative adversarial networks.
\newblock {\em arXiv preprint arXiv:1511.06434}, 2015.

\bibitem{salimans2016improved}
Tim Salimans, Ian Goodfellow, Wojciech Zaremba, Vicki Cheung, Alec Radford, and Xi~Chen.
\newblock Improved techniques for training gans.
\newblock {\em Advances in neural information processing systems}, 29, 2016.

\bibitem{kingma2014semi}
Durk~P Kingma, Shakir Mohamed, Danilo Jimenez~Rezende, and Max Welling.
\newblock Semi-supervised learning with deep generative models.
\newblock {\em Advances in neural information processing systems}, 27, 2014.

\bibitem{goodfellow2014generative}
Ian Goodfellow, Jean Pouget-Abadie, Mehdi Mirza, Bing Xu, David Warde-Farley, Sherjil Ozair, Aaron Courville, and Yoshua Bengio.
\newblock Generative adversarial nets.
\newblock {\em Advances in neural information processing systems}, 27, 2014.

\bibitem{kingma2013auto}
Diederik~P Kingma and Max Welling.
\newblock Auto-encoding variational bayes.
\newblock {\em arXiv preprint arXiv:1312.6114}, 2013.

\bibitem{he2010laplacian}
Xiaofei He, Deng Cai, Yuanlong Shao, Hujun Bao, and Jiawei Han.
\newblock Laplacian regularized gaussian mixture model for data clustering.
\newblock {\em IEEE transactions on knowledge and data engineering}, 23(9):1406--1418, 2010.

\bibitem{gepperth2021gradient}
Alexander Gepperth and Benedikt Pf{\"u}lb.
\newblock Gradient-based training of gaussian mixture models for high-dimensional streaming data.
\newblock {\em Neural Processing Letters}, 53(6):4331--4348, 2021.

\bibitem{papadopoulos2019prosstt}
Nikolaos Papadopoulos, Parra~R Gonzalo, and Johannes S{\"o}ding.
\newblock Prosstt: probabilistic simulation of single-cell rna-seq data for complex differentiation processes.
\newblock {\em Bioinformatics}, 35(18):3517--3519, 2019.

\bibitem{paul2015transcriptional}
Franziska Paul, Ya'ara Arkin, Amir Giladi, Diego~Adhemar Jaitin, Ephraim Kenigsberg, Hadas Keren~Shaul, Deborah Winter, David Lara~Astiaso, Meital Gury, Assaf Weiner, et~al.
\newblock Transcriptional heterogeneity and lineage commitment in myeloid progenitors.
\newblock {\em Cell}, 163(7):1663--1677, 2015.

\bibitem{plass2018cell}
Mireya Plass, Jordi Solana, F~Alexander Wolf, Salah Ayoub, Aristotelis Misios, Petar Gla{\v{z}}ar, Benedikt Obermayer, Fabian~J Theis, Christine Kocks, and Nikolaus Rajewsky.
\newblock Cell type atlas and lineage tree of a whole complex animal by single-cell transcriptomics.
\newblock {\em Science}, 360(6391):eaaq1723, 2018.

\bibitem{wolf2019paga}
F~Alexander Wolf, Fiona~K Hamey, Mireya Plass, Jordi Solana, Joakim~S Dahlin, Berthold G{\"o}ttgens, Nikolaus Rajewsky, Lukas Simon, and Fabian~J Theis.
\newblock Paga: graph abstraction reconciles clustering with trajectory inference through a topology preserving map of single cells.
\newblock {\em Genome biology}, 20:1--9, 2019.

\bibitem{weiskopf2017myeloid}
Kipp Weiskopf, Peter~J Schnorr, Wendy~W Pang, Mark~P Chao, Akanksha Chhabra, Jun Seita, Mingye Feng, and Irving~L Weissman.
\newblock Myeloid cell origins, differentiation, and clinical implications.
\newblock {\em Myeloid Cells in Health and Disease: A Synthesis}, pages 857--875, 2017.

\bibitem{spangrude1988purification}
Gerald~J Spangrude, Shelly Heimfeld, and Irving~L Weissman.
\newblock Purification and characterization of mouse hematopoietic stem cells.
\newblock {\em Science}, 241(4861):58--62, 1988.

\bibitem{vaze2022generalized}
Sagar Vaze, Kai Han, Andrea Vedaldi, and Andrew Zisserman.
\newblock Generalized category discovery.
\newblock In {\em Proceedings of the IEEE/CVF Conference on Computer Vision and Pattern Recognition}, pages 7492--7501, 2022.

\end{thebibliography}

\end{document}